\documentclass{article}

\usepackage{arxiv}

\usepackage[utf8]{inputenc} 
\usepackage[T1]{fontenc}    
\usepackage{hyperref}       
\usepackage{url}            
\usepackage{booktabs}       
\usepackage{amsfonts}       
\usepackage{nicefrac}       
\usepackage{microtype}      
\usepackage{lipsum}		
\usepackage{graphicx}
\def\BibTeX{{\rm B\kern-.05em{\sc i\kern-.025em b}\kern-.08em
    T\kern-.1667em\lower.7ex\hbox{E}\kern-.125emX}}
\usepackage{doi}
\usepackage{tabularx}
\usepackage[table]{xcolor}

\usepackage[nohyperlinks, printonlyused, nolist]{acronym}

\begin{acronym}
    \acro{iot}[IoT]{Internet of Things}
    \acro{mno}[MNO]{Mobile Network Operator}
    \acro{snr}[SNR]{Signal to Noise Ratio}
    \acro{ofdm}[OFDM]{Orthogonal Frequency-Division Multiplexing}
    \acro{plmn}[PLMN]{Public Land Mobile Network}
    \acro{3gpp}[3GPP]{3rd Generation Partnership Project}
    \acro{ce}[CE]{Control Elements}
    \acro{urllc}[uRLLC]{Ultra Reliable and Low Latency Communications}
    \acro{cpps}[CPPS]{cyber-physical production systems}
    \acro{qos}[QoS]{Quality of Service}
    \acro{nm}[NetEm]{network emulation}
    \acro{isac}[ISAC]{Integrated Sensing and Communication}
    \acro{npn}[NPN]{Non-Public Network}
    \acro{ru}[RU]{Radio Unit}
    \acro{du}[DU]{Distributed Unit}
    \acro{5g}[5G]{fifth-generation}
    \acro{6g}[6G]{sixth-generation}
    \acro{ran}[RAN]{Radio Access Network}
    \acro{gnb}[gNB]{Next Generation NodeB}
    \acro{o-cu}[O-CU]{O-RAN Central Unit}
    \acro{o-du}[O-DU]{O-RAN Distributed Unit}
     \acro{o-ru}[O-RU]{O-RAN Radio Unit}
    \acro{oran}[OpenRAN]{Open Radio Access Network}
    \acro{ric}[RIC]{RAN Intelligent Controller}
    \acro{ue}[UE]{User Equipment}
    \acro{plmn}[PLMN]{Public Land Mobile Network}
    \acro{mimo}[MIMO]{Multiple-Input Multiple-Output}
    \acro{ng-ran}[NG-RAN]{Next Generation Radio Access Network}
    \acro{papr}[PAPR]{Peak-to-Average Power Ratio}
    \acro{rnti}[RNTI]{Radio Network Temporary Identifier}
    \acro{dmrs}[DMRS]{Demodulation Reference Signal}
    \acro{pci}[PCI]{Physical Cell Identifier}
    \acro{csirs}[CSI-RS]{Channel State Information - Reference Signal}
    \acro{mitm}[MITM]{Man-In-The-Middle}
    \acro{near-rt ric}[near-RT RIC]{Near-Real-Time RAN Intelligent Controller}
    \acro{non-rt ric}[non-RT RIC]{Non-Real-Time RAN Intelligent Controller}
    \acro{e2}[E2]{E2 Interface (RIC-DU/CU control and telemetry interface)}
    \acro{o1}[O1]{O1 Interface (Management interface)}
    \acro{prb}[PRB]{Physical Resource Block}
    \acro{jcas}[JCAS]{Joint Communication and Sensing}
    \acro{embb}[eMBB]{Enhanced Mobile Broadband}
    \acro{ai}[AI]{Artificial Intelligence}
    \acro{ml}[ML]{Machine Learning}
    \acro{c-ran}[C-RAN]{Centralized Radio Access Network}
    \acro{o-cloud}[O-Cloud]{O-RAN Cloud Infrastructure}
    \acro{smo}[SMO]{Service Management and Orchestration}
    \acro{rrc}[RRC]{Radio Resource Control}
    \acro{sdap}[SDAP]{Service Data Adaptation Protocol}
    \acro{pdcp}[PDCP]{Packet Data Convergence Protocol}
    \acro{rlc}[RLC]{Radio Link Control}
    \acro{rf}[RF]{Radio Frequency}
    \acro{mac}[MAC]{Medium Access Control}
    \acro{phy}[PHY]{Physical Layer}
    \acro{scs}[SCS]{Sub Carrier Spacing}
    \acro{tdd}[TDD]{Time-Divison-Duplex}
    \acro{fdd}[FDD]{Frequency Division Duplex}
    \acro{bwp}[BWP]{Bandwidth Parts}
    \acro{uav}[UAV]{Unmanned Aerial Vehicle}
    \acro{eu}[EU]{European Union}
    \acro{fft}[FFT]{Fast-Fourier Transformation}
    \acro{ulpi}[ULPI]{Uplink Performance Improvement}
    \acro{fmcw}[FMCW]{Frequency Modulated Continuous Wave}
    \acro{otfs}[OTFS]{Orthogonal Time Frequency Space}
    \acro{dft}[DFT]{Discrete Fourier Transform}
    \acro{gdpr}[GDPR]{General Data Protection Regulation}
    \acro{nearric}[Near-RT RIC]{Near-Real-Time RIC}
    \acro{sm}[SM]{Slice Management}
    \acro{sib}[SIB]{System Information Block}
    \acro{cots}[COTS]{Commercial Off-The-Shelf}
    \acro{agv}[AGV]{Automated Guided Vehicle}
    \acro{iq}[I/Q]{In-phase and Quadrature}
    \acro{upf}[UPF]{User Plane Function}
   \acro{rnti}[RNTI]{Radio Network Temporary Identifier}
\end{acronym}

\title{Analysis of an Architecture for Integrated Sensing and Communication in 5G OpenRAN}

\author{ {Daniel~Lindenschmitt}\\
	Institute for Wireless Communication \\and Navigation\\
	RPTU Kaiserslautern-Landau\\
	\texttt{daniel.lindenschmitt@rptu.de} \\
	\And
	{Tobias~Jung} \\
    Institute for Smart Electronics and Systems\\
    FAU Erlangen-Nuremberg\\
	\texttt{tobias.jung@fau.de} \\
    \And
	{Prudhvi~Kumar~Kakani} \\
	Institute for Wireless Communication \\and Navigation\\
	RPTU Kaiserslautern-Landau\\
	\texttt{prudhvi.kakani@rptu.de} \\
     \And
	{Torsten~Reissland} \\
    Institute for Smart Electronics and Systems\\
    FAU Erlangen-Nuremberg\\
	\texttt{torsten.reissland@fau.de} \\
 	\And
	{Norman~Franchi} \\
    Institute for Smart Electronics and Systems\\
    FAU Erlangen-Nuremberg\\
	\texttt{norman.franchi@fau.de} \\
 \And
	{Hans D.~Schotten}\\
	Institute for Wireless Communication \\and Navigation\\
	RPTU Kaiserslautern-Landau\\
	\texttt{schotten@rptu.de} \
}

\date{}

\renewcommand{\shorttitle}{\textit{arXiv} Template}

\begin{document}
\maketitle

\begin{abstract}
This paper analyzes the functional requirements and architectural considerations for \ac{isac} in a 5G \ac{oran} environment, with emphasis on secure and modular deployment. Focusing on a mono-static, half-duplex sensing approach, it evaluates radar setup options, signal types, and processing placement within the \ac{ran}, considering performance and security implications. The proposed architecture minimizes hardware modifications by leveraging sniffer \acp{ru} and existing \ac{oran} fronthaul interfaces, while protecting sensitive \ac{iq} data and control traffic against potential attacks. Security threats, such as passive sensing, spoofing, and privacy violations, are mapped to mitigation strategies within the \ac{oran} framework. The result is a deployment blueprint applicable to both \acp{plmn} and \acp{npn}, supporting future 6G \ac{isac} capabilities in a standards-compliant manner.
\end{abstract}


\section{Introduction}
\label{sec:1_introduction}
Using wireless networks to both communicate and sense the environment is shaping up to be a key feature of future 6G systems. \ac{isac} reuses existing infrastructure to identify problems with equipment at an early stage, tracking movement in real-time, or getting a clearer picture of what’s happening around the network. Despite its potential, \ac{isac} integration remains technically challenging under the current cellular network architectures. \ac{oran} provides a flexible and standardized architecture that disaggregates traditional base station functions into modular components. With its open interfaces, virtualized components, and inclusion of with the \ac{ric} through xApps and rApps, \ac{oran} is a promising platform for implementing advanced features like \ac{isac}. Particularly relevant are its support for network slicing, programmable scheduling, and distributed processing capabilities.

In this paper, we examine the architectural modifications and functional extensions needed to enable \ac{isac} within a 5G-based \ac{oran} system. We focus on sensing configurations, signal processing requirements, data security, and slice coexistence. The results aim to guide the design of scalable \ac{isac}-capable RAN deployments, helping to pave the way toward practical 6G applications.

\section{Background}
\label{sec:rel_work}
\subsection{\ac{isac}}
\label{subsec:isac}
\ac{isac} is an emerging technology which, in the context of mobile communications, allows the reuse of communication resources to enable different degrees of sensing capabilities. By now, numerous use cases for \ac{isac} were identified and their requirements derived. 
Regarding the architecture of the sensing functionality, several options are possible, each with their respective advantages and drawbacks. A mono-static architecture, in our definition, denotes a sensing system in which the same \ac{ru} is responsible for the transmission and reception of the signals. The transmitting and receiving antennas can be identical, but don't have to be. Such a system avoids any synchronization impairments. Also, one has full control over the antenna placements, which offers the possibility of optimizing the beam pattern. The main drawback is the need for an \ac{ru} which is capable of full-duplex operation, increasing the complexity of the \ac{ru}, mainly due to the required isolation between receiver and transmitter \cite{faghih_naini2024}. Such \acp{ru} are, as of today, not commercially available.
The second option would be co-located bistatic \acp{ru}. Achieving sufficient isolation between transmitter and receiver is easier in this case, also synchronization through a common local oscillator is possible. One major drawback of this approach is the significantly more expensive hardware for two \acp{ru}. A distributed \ac{isac} could consist of either multiple \acp{ru} on different sites or an \ac{ru} and one or more \acp{ue}. However, the latter approach is not in the scope of this paper. The main drawback of using distributed \acp{ru} is the limitation on areas which are covered by at least two of them. As this limits the applicability, we will focus on the co-located setup.


\subsection{OpenRAN}
\label{subsec:openran}
The \ac{oran} architecture provides a flexible and modular framework that is well-suited for integrating sensing functionality into mobile networks. In contrast to traditional monolithic \ac{ran} designs, \ac{oran} disaggregates the \ac{ran} into functional elements such as the \ac{o-ru}, \ac{o-du}, and \ac{o-cu}, connected via open, standardized interfaces (e.g., 7.2x for fronthaul, \ac{e2} for \ac{ric} interaction) \cite{alam2024comprehensivetutorialsurveyoran, oranarch2022, Asif_Trust_OpenRAN}. This division enables a modular approach to implement \ac{isac} functions without the need for system-wide hardware redesign. For example, radar processing modules can be integrated into the \ac{o-ru} for local sensing or into the \ac{o-du} for centralized processing. Additionally, \ac{oran} enables network programmability via the \ac{ran} Intelligent Controller (\ac{ric}), allowing near-real-time optimization of radio resources and potentially hosting \ac{isac} control logic in the form of xApps and rApps \cite{doro2022dapps}. Another key advantage of \ac{oran} is its support for network slicing, which enables the allocation of customizable radio resources to individual slices. This feature is particularly valuable when \ac{isac} and communication services need to operate concurrently, as it allows partitioning of physical-layer parameters (e.g., subcarrier spacing, \ac{tdd} configurations) to match the specific requirements of each service \cite{habibi2024aimldrivensmoframeworkoran, alam2024comprehensivetutorialsurveyoran}. By leveraging these architectural flexibilities, \ac{oran} facilitates the deployment of scalable and adaptable \ac{isac} functionalities.

\subsection{\acp{npn}}
\label{subsec:npn}
The emergence of \acp{npn} in 5G reflects a shift towards localized and purpose-specific mobile communication systems. Unlike traditional \acp{plmn}, \acp{npn} are built and managed for individual operators to serve dedicated needs, whether in manufacturing, transport, or critical infrastructure. Their inclusion in the \ac{3gpp} specification, particularly in Release 16, formalized two main deployment models: standalone private networks and hybrid private networks, where parts of the network are integrated in the infrastructure of \acp{mno}. Several studies have explored the design choices and limitations associated with deploying \acp{npn}. One key aspect is the limited spectrum and how it is allocated and licensed at a national level.  Matinmikko-Blue et al. discussed, how local licensing schemes vary across countries~\cite{Spectrum}. The introduced approaches enable customized services, but the authors also raise questions about interoperability and regulation, especially when networks are mobile or temporary in nature. For example, Lindenschmitt et al. discuss how Nomadic \acp{npn} can extend service coverage to locations lacking permanent infrastructure~\cite{lindenschmitt2024usecases, towardsnomadic}. From a systems perspective, architectural flexibility is essential when adapting mobile networks to serve enterprise or localized use cases. 
Another trend influencing \ac{npn} development is the rise of \ac{oran} architectures. By decoupling hardware and software, and standardizing interfaces, \ac{oran} enables more flexible implementations. Some recent efforts within the O-RAN Alliance have explored how custom \ac{ric} applications might be used to manage and optimize private networks~\cite{oranarch2022}.

\section{Secure Implementation of ISAC in the 5G O-RAN Architecture}
\label{sec:3_implementation}

During the analysis of integrating network side based radar sensing within the 5G \ac{oran} architecture, we first conducted a preliminary analysis to identify associated key challenges. Then, by examining these challenges in relation to fundamental system design decisions, we mapped correlations between challenges and system design decisions to break down complexity and allow focused analysis of identified major subsystems. In this section each of them will be introduced and have related works and challenges highlighted. 

\subsection{Physical Radar Setup}
\label{subsec:radar_setup_3}
\begin{figure}[ht]
    \centering
    \includegraphics[width=1\linewidth]{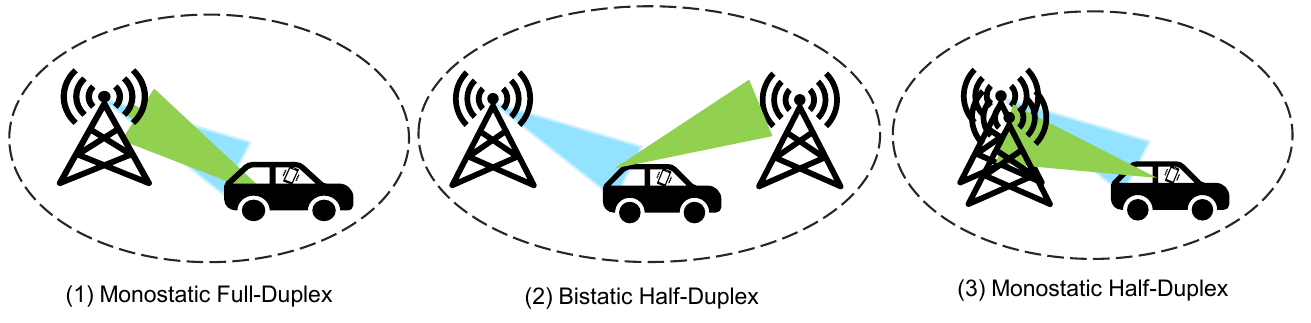}
    \caption{Possible physical radar setups for 6G \ac{isac}}
    \label{fig:radar_setup}
\end{figure}

The basic requirement for sensing the environment is a simultaneous capture of the reflections of transmitted radio waves. For mobile communications in higher frequency bands, and subsequently larger bandwidth, \ac{tdd} transmission schemes are most commonly found. Consequently, there has been no need for full-duplex base stations. For our analysis, three possible implementations of radar systems are considered, as illustrated in Fig. \ref{fig:radar_setup}.

The first possible implementation, which was first discussed by Nokia Bell Labs' Wild et al. \cite{wild_2021,wild20236gintegratedsensingcommunication}, is to use a second half-duplex capable \ac{ru} for the sole task of detecting reflected \ac{rf} waves. Both half-duplex capable \acp{ru} are then co-located with a slight spatial offset, building a mono-static radar system. This significantly reduces the problem of self-interference between the two radios in comparison to a singular full-duplex mono-static radio \cite{wild20236gintegratedsensingcommunication}. Beyond reducing self-interference, the system could be further enhanced by leveraging the second \ac{ru} to increase the \ac{snr} of uplink communication. Moreover, it offers flexibility during deployment allowing to expand pure communication systems into \ac{isac} ones through simple integration of an extra \ac{ru}. In a similar sense a switch from mono to bistatic radar, see also option 2 in Fig. \ref{fig:radar_setup}, would be possible. Bi- or multi-static radar systems can balance higher infrastructural and technical requirements against better sensing performance. On a further note, distributed \acp{ru} can ease access to full duplex \ac{mimo} communication technologies in the same frequency band \cite{alexandropoulos2022duplexmassivemimoarchitectures} since self-interference is reduced due to physical locality. Finally, a third implementation option would involve developing a full duplex capable radio unit. This is challenging due to self-interference, especially regarding the high transmission power of the base station comparatively to the incoming power of \ac{ue} signals, as well as the increasing number of antenna elements. 

\subsection{Signal data}
\label{subsec:signal_data_3}
This subset of the system design encompasses data types used for transmission, including derived overarching physical design parameters. Three data types are discussed: Communication user data, communication reference signals and pilot sensing signals. As the basis for our analysis, we assume \ac{ofdm} as the fundamental \ac{isac} waveform, while noting that the analysis can be similarly applied to other waveforms such as \ac{otfs} or DFT-S \ac{ofdm}.

Sensing with \ac{ofdm} waveforms is based on using the channel information for range and velocity resolution \cite{Braun2014_1000038892}. This allows the usage of user data for sensing without a significant loss of sensing performance. The general challenge of \ac{ofdm} sensing stems from its high \ac{papr}, though this can be reduced through signal conditioning or other waveform design techniques \cite{wild_2021}. If the \ac{isac} system is using random user data instead of deterministic reference or pilot signals, malicious, non-cooperative entities are hindered in manipulating the resulting sensing positions of real or fake targets by sending tampered waveforms overshadowing the reflected signal \cite{10149610}. Furthermore, since sensing performance correlates to available time and frequency resources, user data can be sent repeatedly or with lower modulation orders to not only increase resolution but also communication reliability and latency.

Instead of using optimized pilots for sensing, communication reference signals could be used. However, compared to the total amount of available time and frequency resources reference signals, especially for higher bandwidths, only occupy a marginal part. Consequently, these resources are, under consideration of the current standard, not enough by themselves to fulfill the majority of sensing applications. Considering malicious entities these signals are also more challenging to decode because some rely on the necessity of prior knowledge of the cell side such as \ac{rnti} or \acp{pci}, though the possibility to spoof these also exists \cite{ludant2024unprotected4g5gcontrolprocedures}. This includes especially the more frequently occurring reference signals, such as \ac{dmrs} or \ac{csirs} in 5G, though the associate risk should be researched more thoroughly including possible protection against \ac{rnti} sniffing.

As for the third option, a higher impact in terms of passive sniffer and target manipulation vulnerability, but also conversely in terms of sensing performance, is achieved by using additional sensing based waveforms, e.g. \ac{fmcw} or other pilot signals like Zadoff-Chu sequences \cite{ofdm_zadoff}. A base constraint or trade-off for using additional pilot signals for sensing is that frequency and time resources need to be available. Furthermore, this would require additional energy consumption for purely improving sensing quality, which might be counterintuitive to the core of \ac{isac}. Contrary, a benefit of reference and pilot signals, especially relating to a bi- or multi-static radar system, is that they are known at the receiver, enabling radar processing directly without demodulation or other prior knowledge of the transmitted signal.

\subsection{Placement of Radar Processing}
\label{subsec:rad_proc_3}
\renewcommand{\arraystretch}{1.2} 
\begin{table*}[t]
\centering
\caption{Comparison of challenges for radar processing in DUs and RUs}
\rowcolors{2}{blue!5}{white} 
\begin{tabularx}{\textwidth}{|>{\raggedright\arraybackslash}X|>{\raggedright\arraybackslash}X|}
\hline
\rowcolor{blue!30}
\textbf{Distributed Unit} & \textbf{Radio Unit} \\
\hline
\textbf{Fronthaul Load:} As a baseline, increases for both cases because the captured transmitted waveform must also be processed. But additionally, for this case, \ac{iq} compression methods on the receiver side can only be used to a limited extent, due to distortions of the channel information. &
\textbf{RU/DU Architecture Adjustment:} Required, as layer demapping and channel estimation must be performed before radar processing. This implies shifting these functions from the DU to the RU or duplicating them. \\
\hline
\textbf{Fronthaul Data Security:} \ac{iq} data must be cryptographically protected during fronthault transmission, as they carry privacy and security sensitive information. &
\textbf{Radar Tasks:} The RU must perform the initial radar resolution (creation of range-doppler maps), which must be cryptographically protected before being sent to the DU. \\
\hline
\mbox{} &
\textbf{Synchronization and Transmitted Waveform:} The Transmitting and Receiving RUs need to be synchronized including knowledge of the transmitted waveform at receiver side. \\
\hline
\end{tabularx}
\label{tab:ru_vs_du}
\end{table*}

A major point of discussion is the placement of radar processing functions in the \ac{ran}. For \ac{oran} the two most likely components for the first step of radar processing include the \ac{ru} and \ac{du}, or a \ac{ric} that is connected to the \ac{du}. An overview of a comparison when integrating \ac{isac} into the \ac{oran}architecture for the placement of radar processing is given in Table \ref{tab:ru_vs_du}. Of note is that a radar processing in the \ac{ru} requires either a \ac{ru} with an adjusted architecture or a duplication of some \ac{du} functions into the \ac{ru}. Interestingly, for the case of improving extreme \ac{mimo} uplink similar changes have  been discussed for the category B fronthaul \cite{Ericsson2025DrivingOpenRAN}. The in \cite{Ericsson2025DrivingOpenRAN} discussed category B \ac{ulpi} fronthaul places channel estimation in the \ac{ru} fitting well with the architectural \ac{isac} requirement. Another interesting challenge is that the \ac{iq} symbols at the receiver side need to be protected to disallow \ac{mitm} tampering or sniffer attacks on the fronthaul transport to gain information about the environment. Other challenges, especially for multi-static setups, include the time synchronization and transmitted \ac{iq} symbols for channel estimation and subsequent radar sensing in the \acp{ru}.

\section{Evaluation of Subsystems}
\label{sec:4_evaluation}

This chapter evaluates the introduced subsystems of section \ref{sec:3_implementation}. Open challenges and optimizations for further research are discussed in the last subsection.

\subsection{Deployment of Radar Setup}
\label{subsec:radar_setup_4}
\begin{table*}[t]
\centering
\caption{Comparison of radar deployment configurations}
\rowcolors{2}{blue!5}{white}
\begin{tabularx}{\textwidth}{|>{\raggedright\arraybackslash}p{3.5cm}
              |>{\raggedright\arraybackslash}X
              |>{\raggedright\arraybackslash}X
              |>{\raggedright\arraybackslash}X|}
\hline
\rowcolor{blue!30}
\textbf{Aspect} & \textbf{Mono-static \& Half-Duplex} & \textbf{Mono-static \& Full-Duplex} & \textbf{Bi-/Multi-static \& Half-Duplex} \\
\hline
Self-interference & + Low & - High & ++ Not present \\
\hline
Hardware Modifications Required & + Minimal & - RU replacement & + Minimal \\
\hline
Infrastructure Requirements for Sensing & - second antenna \& cabling needed & + None & -- Additional antenna site \& Second antenna \& cabling needed \\
\hline
Impact on Communication & + Second RU improves upload & - None & + Second RU improves upload \\
\hline
Synchronization & + Simple & + Simple & - Complex \\
\hline
Energy Efficiency & - Two RUs & + One RU & - Two RUs \\
\hline
\end{tabularx}
\label{tab:radar_comparison}
\end{table*}

\subsubsection{Performance}
This section presents a comparison of various radar system configurations and their impact on overall system performance. The overview is given in Table \ref{tab:radar_comparison}, which compares the influence of different radar systems depending on specified aspects relevant for the overall system performance. Apart from the evaluations in the table, it should be noted that the choice of the radar system itself does not have a direct impact on the sensing performance. Specifically, only the covered area and the signal-to-noise ratio of the received signal influence the performance, which is assumed to be similar for all discussed radar systems.

\subsubsection{Security}
The security of radar deployment configurations in \ac{oran}  \ac{isac} systems depends primarily on the protection of the fronthaul and physical access points. While mono-static configurations simplify access control through co-location, multi-static systems introduce broader coverage at the cost of increased physical exposure and synchronization complexity. However, the most critical vulnerability lies in the unencrypted \ac{iq} data transmitted over the fronthaul, which is vulnerable to eavesdropping, spoofing, and synchronization attacks. Since \ac{3gpp} does not mandate fronthaul encryption, a single compromised link can undermine an entire base station. Effective countermeasures include encrypting radar data at the \ac{ru}, hardening synchronization protocols, and adopting tamper-resistant hardware strategies for distributed \acp{ru}.

\subsection{Signal data}
\label{subsec:signal_data_4}

\subsubsection{Performance}
Generally the performance of the data types is simple: Sensing optimized pilot signals have the best sensing performance, but add nothing towards communication, while user data and reference signals are necessary for communications but have more limited sensing performance. Due to security reason, see below, stochastic user data should be used. For our further work we assume that the integrity and information disclosure due to sniffers warrants the use of stochastic user data as much as possible. Though, the exact amount of stochastic to deterministic data or if detection algorithms are enough to authenticate reflected waveforms needs to be further investigated.

\subsubsection{Security}
The type of signal used for sensing directly affects the system’s resilience to passive and active attacks. Stochastic user data offers the best defense, as its unpredictability hinders adversarial channel estimation and false target injection. In contrast, deterministic pilot signals—though superior for sensing accuracy are highly exploitable for spoofing and passive radar. Communication reference signals occupy a middle ground but remain partially vulnerable even if enhanced by masking techniques like  \ac{rnti} scrambling. To safeguard \ac{isac} operations, it is recommended to prioritize stochastic or randomized waveforms, avoid pilot signals in public networks, and implement \ac{phy}-layer signal authentication to counter spoofing and signal replay.

\subsection{Placement of Radar Processing}
\label{subsec:rad_proc_4}

\subsubsection{Performance}
The main differences in radar processing placements include real-time capability, fronthaul load and cost efficiency. The latter arising from the fact that processing in \acp{du} or higher components is cheaper since the calculations are done on \ac{cots} server hardware. However, processing in the \ac{ru} lessens the load on the fronthaul, since only compressed \ac{iq} data of the uplink resource elements need to transported. As for real-time capability: While initial radar processing such as the general and basic transformations of \ac{iq} samples into range-doppler maps can be done in the \ac{ru}, the algorithms for object detection, classification, sensing applications or possible anonymization and/or sensing control would increase the complexity of \ac{ru} development and deployment. 

\subsubsection{Security}
Choosing where to perform radar signal processing has major implications for confidentiality and system integrity. Processing at the \ac{ru} minimizes fronthaul exposure by converting raw \ac{iq} data into encrypted range-doppler maps before transmission, thereby mitigating man-in-the-middle and eavesdropping threats. However, this shifts trust to the edge and may require Trusted Execution Environments. \ac{du}-based processing benefits from centralized security controls but requires full cryptographic protection of sensitive fronthaul data. Given comparable computational costs, the recommended approach is early \ac{ru}-side processing with \ac{iq} scrambling or compression, combined with a zero-trust \ac{ru}–DU communication framework for robust and scalable \ac{isac} deployment.

\subsection{Open Research}
\label{subsec:open_research}
By evaluating the defined sub-systems on radar setup, signal data and radar processing important questions towards the integration on \ac{isac} into the \ac{oran} architecture have been successfully addressed, while others remain unanswered and need further investigation. One challenge is how \ac{isac} can be integrated into sliced networks. Slicing allows the network to assign different physical layer settings—like subcarrier spacing or TDD patterns—to different use cases. However, it’s still not well understood how these variations affect sensing quality or how schedulers should react to balance sensing and communication needs. Besides that, valuable implementation of \ac{isac} functionalities into \acp{npn} needs further research, especially in the domain of nomadic, where spectrum allocation and usage need to be defined with appropriate lead time, or in self-organizing, organic networks~\cite{organic6g}. In addition the integration of sensing capabilities into the existing network-management of \ac{oran} based \acp{npn} is unclear. There’s also little agreement on how to standardize the handling of sensing data or how sensing tasks should be configured and controlled within the network.
Finally, if multiple radio units are involved in sensing, as in multi-static setups, coordination and timing become critical, and solutions for this are still immature. 

\section{Architectural Recommendation}
\label{sec:5_architecture}
\begin{figure*}[t]
    \centering
    \includegraphics[width=1.0\textwidth]{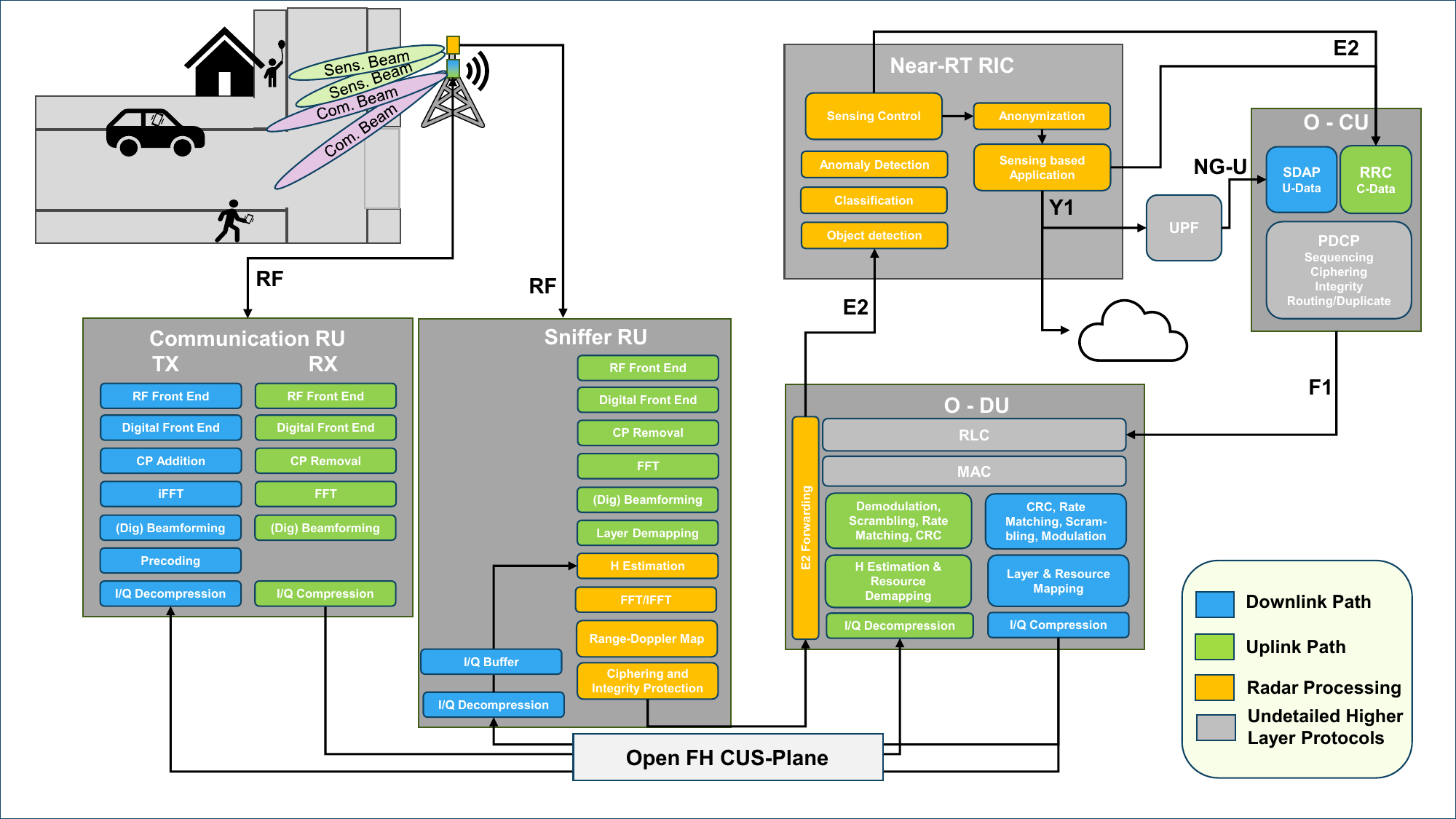}
    \caption{Recommended 5G OpenRAN architecture for \ac{isac}}
    \label{fig:architecture_recommendation}
\end{figure*}


This section presents our recommendation for the integration of \ac{isac} in the 5G \ac{oran} architecture, that can be directly integrated into \ac{plmn} and \acp{npn} maximizing value services for industrial deployments. Our motivation lies in the fact that we see \ac{isac} as a key enabler for a multitude of use cases \cite{inproceedings2}. For ease of deployment it is necessary that no standardization is required, and only minimal changes are needed. For usage across different real-time capabilities and even for critical use cases we want to ensure that security and privacy is integrated at the base of our design.

Figure \ref {fig:architecture_recommendation} shows a slightly expanded overview of the specified Category B Fronthaul of the \ac{oran} architecture. See section \ref{subsec:openran} for an explanation on the base architecture. On the top left side a common city scenario is illustrated in conjunction with a mono-static half-duplex setup with two \acp{ru}, one for communication including transmission and reception, and one simply for sensing purposes. In the rest of the Figure the blue and green functions represent the downlink and uplink path respectively, whereas the greyed functions are general protocols on layer 2/3 that are not subdivided into functions or either link direction. Finally, in orange are the additional radar functions that are necessary to securely integrate radar sensing in 5G \ac{oran}.

We chose a mono-static half-duplex radar system mainly due to the early possibility of implementation in comparison to full-duplex system, as well as enabling the complete reuse of the current infrastructure and radios. Basically, the communication \ac{ru}’s downlink and uplink path is completely untouched in regards to the current state of Category B \ac{oran} fronthaul necessitating no standardization or specification changes. The only addition would be that the sniffer \ac{ru} has to be synchronized to the same \ac{du} and receive the same compressed \ac{iq} samples. Thusly the economic prospects are twofold without any loss of security or communication performance. First radar functions can be added as necessary to such 5G \ac{oran} cell sites by integrating a sniffer \ac{ru} as introduced in \cite{wild20236gintegratedsensingcommunication}. Secondly required infrastructure changes are minor, e.g. compared to a bi-static setup.

Due to security and privacy concerns we recommend that sensing in general will only be done through the usage of stochastic user data. This stems due to threats of tampering with reflected waveforms as well as passive sniffing discussed in section \ref{subsec:signal_data_3}. Similar threats can be applied to the \ac{iq} data transmitted over the Open FH CUS-Plane, between \acp{ru} and \ac{du} in Figure \ref{fig:architecture_recommendation}. After a comparison of the necessary amount of computational power between initial radar processing and cryptographic protection of high data rates of \ac{iq} fronthaul data, our results showed a similar order of compute resource cost. Consequently, we decided to execute the initial radar processing in the sniffer \ac{ru} and simply encrypt and integrity protect the resulting range-doppler maps. This still leaves the protection of \ac{iq} data of the communication \ac{ru} in the uplink direction to the \ac{du}. For these it is planned to use high \ac{iq} compression to reduce the sniffing capability of any attacker on the fronthaul drastically by degrading the inherent channel information. The exact parametrization and algorithm still needs to be researched in detail.


For current \ac{oran} the necessary radar functions for further processing of range-doppler maps should be added to the \ac{nearric} or in the future in the form of real time capable dApps. These radar functions include in the processing order: Object detection, classification, anomaly detection and later on anonymization. The last one especially is to anonymize any sensing information that might infringe on regulations such as the \ac{gdpr} in the \ac{eu}. We can derive the  state and quality of sensing from the results of object detection, classification and anomaly detection in the “Sensing Control” function. This way security problems are recognized early on and the current quality of sensing can be compared to the targeted one which was defined by the corresponding applications. If tampering is detected or discrepancies occur in the targeted quality, control commands can be sent over the E2 interface towards the \ac{rrc} to adapt network configurations. Some possible configurations include, \ac{bwp}, \ac{tdd} patterns or beam steering. To be clear the latter refers to the steering of the sensing beams of the sniffer \ac{ru}. For the current specification this could be done by a reinterpretation from the sniffer \ac{ru} of the \textit{ssb-PositionsInBurst} parameter in the \textit{ServingCellConfigCommonSIB} message \cite{etsi_ts138331_v18_6_0}. Since the sniffer is not transmitting anything these protocols could be used for first deployments though the bitmap is limited to 64 beam options and does not include a granular adjustment of beam weights.

Afterwards, the privacy protected sensing results are forwarded and processed in “Sensing Application”. In Figure \ref{fig:architecture_recommendation} this block is placed in the \ac{near-rt ric} to enable fast response times by leveraging the near real-time control loop of \ac{oran}, but the necessary quality of sensing, including real-time capability, largely depends on the application \cite{6G_pers}, \cite{isac_open6ghub}. For example, in the case of steering an \ac{agv}, one possible category of sensing applications include a forwarding of steering data from a sensing application deployed in the \ac{near-rt ric} towards the \ac{ue}, possibly sending the information over the Y1 interface and the \ac{upf}. Enabling the same use case can also be done by placing the same sensing application in the \ac{ue} and transmitting the anonymized sensing information directly to the \ac{ue} over the same path. On the other hand, for non real-time based applications long-time environmental results could be stored into a cloud. Finally, the network itself can profit from sensing information through the deployment of inherent network-based application such as beam steering or mobility management by leveraging knowledge about the environment. To integrate such network-based applications like predictive beam steering for the communication \ac{ru} a different protocol than for the sniffer \ac{ru} would be preferred since the adjustment would be specific to individual \acp{ue}. After some initial research we recommend to use \textit{RRCConnectionReconfiguration} protocol which is dedicated for connection management of singular \acp{ue} and allows for \textit{dedicatedSIB1-Delivery} which in turn includes \textit{ServingCellConfigCommonSIB} again \cite{etsi_ts138331_v18_6_0}. There are cases where faster beam steering is necessary. However, the reasons this approach was chosen are manifold. First, in the current specifications there are no protocols for beam adjustments of lower layers from the network side. Admittedly there does exist a lower layer protocol (beam failure management) that allows beam steering before layer 3, but it is initiated by the \ac{ue}. Additionally, at the current state of the standardization most control commands should be propagated to the lower layer through the CU due to the security functions added by the \ac{pdcp}, contrary to L2 \ac{mac} \acp{ce} that are shown to be security sensitive \cite{ludant2024unprotected4g5gcontrolprocedures}. On top of that \acp{ric} allow for a more open deployment and development process supporting radar processing and fusion algorithms.


\section{Conclusion}
\label{sec:6_conclusion}
This paper has analyzed the integration of \ac{isac} within a 5G \ac{oran} architecture. We identified and analyzed the key system components that need to be adapted or extended, ranging from radar signal design and data security to the physical setup and processing placement within the RAN. Furthermore, an architecture was proposed, enabling secure and efficient \ac{isac} integration without disrupting existing \acp{oran}. The paper showed that the integration is feasible, especially in the field of mono-static half-duplex setup with limited hardware adaptions. Distributing the radar processing to the \ac{ru} reduce fronthaul load and can in addition improve sensing latency, while still maintaining data protection through compression and encryption techniques.

Future work in this domain needs to analyze the interaction between \ac{isac} and network slicing, particularly when slices carry different waveform or timing configurations, as well as detailed research on reliable \ac{isac} integration into \acp{npn}, e.g. in nomadic or temporal scenarios. A better understanding of how sensing data should be managed across layers, and how to ensure performance and privacy, will be key for turning \ac{isac} into a standard feature of future 6G mobile networks.

\section{Acknowledgment}
The authors acknowledge the financial support by the German \textit{Federal Ministry for Research, Technology and Space (BMFTR)} within the project Open6GHub \{16KISK004,16KISK005\} and 6G-CampuSens \{16KISK206,16KISK208\}.

\bibliographystyle{ieeetr} 
{%
\bibliography{references}
}%
\end{document}